\documentclass{PoS}
\catcode`\@=11
\def\gsim{\ifmmode{\,\mathrel{\mathpalette\@versim>\,}}
    \else{$\,\mathrel{\mathpalette\@versim>}\,$}\fi}
\def\lsim{\ifmmode{\,\mathrel{\mathpalette\@versim<\,}}
    \else{$\,\mathrel{\mathpalette\@versim<}\,$}\fi}
\def\@versim#1#2{\lower 2.9truept \vbox{\baselineskip 0pt \lineskip
    0.5truept \ialign{$\m@th#1\hfil##\hfil$\crcr#2\crcr\sim\crcr}}}
\catcode`\@=12

\def\av#1{\langle#1\rangle}
\def\rms{\rm rms}
\def\yr{{\rm yr}}

\def\fQ{f_{\rm Q}}
\def\fmicroQ{f_{\rm \mu Q}}

\def\Ltwoghzmed{L_{\rm 2\,GHz,med}}
\def\Ltwoghz{L_{\rm 2\,GHz}}
\def\Leightghz{L_{\rm 8\,GHz}}
\def\alphaghz{\alpha_{\rm 2-8\,Hz}}

\bibliography{/path/to/your/bibliographic/file}
\title{Clues from microquasars to the origin of radio-loudness of quasars}

\ShortTitle{Clues from microquasars to the origin of radio-loudness of
  quasars}

\author{\speaker{Carlo Nipoti}\\
        Università di Bologna, Italy\\
        E-mail: \email{carlo.nipoti@unibo.it}}

\author{Katherine M.\ Blundell\\
        University of Oxford, Department of Physics, Keble
  Road, Oxford, OX1 3RH, U.K.\\}

\author{James Binney\\
        University of Oxford, Department of Physics, Keble
  Road, Oxford, OX1 3RH, U.K.\\}

\abstract{We analysed the long-term variability of four microquasars
(GRS 1915+105, Cyg X-1, Cyg X-3, and Sco X-1) in radio and X rays.
The results of our analysis indicate the existence of two distinct
modes of energy output, which we refer to as the `coupled' mode and
the `flaring' mode.  The coupled mode is responsible for mildly
fluctuating, flat-spectrum radio emission, coupled with the X-ray
emission; the flaring mode produces powerful, steep-spectrum radio
flares, with no significant counterpart in X rays.  We find that the
fraction of time spent by a typical microquasar in the flaring mode is
similar to the fraction of quasars that are radio-loud.  This is
consistent with the hypothesis that radio-loudness of quasars is a
function of the epoch at which the source is observed.}

\FullConference{VI Microquasar Workshop: Microquasars and Beyond\\
                 September 18-22, 2006\\
                 Como, Italy}

\begin{document}

\section{Introduction}

Microquasars are powered by accreting stellar mass black holes or
neutron stars, so they can be considered scale models of active
galactic nuclei (AGN). The characteristic times for the evolution of
these sources scale with the mass of the accreting object, so
long-term monitoring of microquasars represents a unique opportunity
to get insights on the variability properties of quasars over
timescales up to $10^8-10^9\yr$. We exploit this scaling between
microquasars and AGN to obtain clues to the long-standing question of
the origin of radio-loudness of quasars. In particular, we explore the
possibility that many quasars that are now radio-quiet may be at times
radio-loud, while currently radio-loud quasars may often be
radio-quiet.

We analysed the long-term variability in radio and in X-rays of the
microquasars GRS\,1915+105, Cyg\,X-3, Sco\,X-1 and Cyg\,X-1. We used
data from the archive of the US National Radio Astronomy Observatory
Green Bank
Interferometer\footnote{http://www.gb.nrao.edu/fgdocs/gbi/gbint.html}
(at 2.25~GHz and 8.3~GHz) and the Rossi X-ray Timing Explorer All-Sky
Monitor\footnote{http://xte.mit.edu/ASM\_lc.html} (at 2-10 keV). In
particular, we considered daily-averaged X-ray and radio
light-curves. The X-ray light-curves of the four sources refer to the
period $50087-53300$ (J.D.$-2400000$); the periods covered by the
radio light curves differ from source to source ($49485-51823$ for
GRS\,1915+105, $44915-51823$ for Cyg\,X-3, $50638-51823$ for Sco\,X-1,
$50412-51823$ for Cyg\,X-1; J.D.$-2400000$). Here we briefly report
some results we presented and discussed in detail in \cite{NBB05},
together with new results on the variability of the radio spectra of
these sources.

\begin{figure}
\centering{ 
\label{figps}
\includegraphics[width=.6\textwidth]{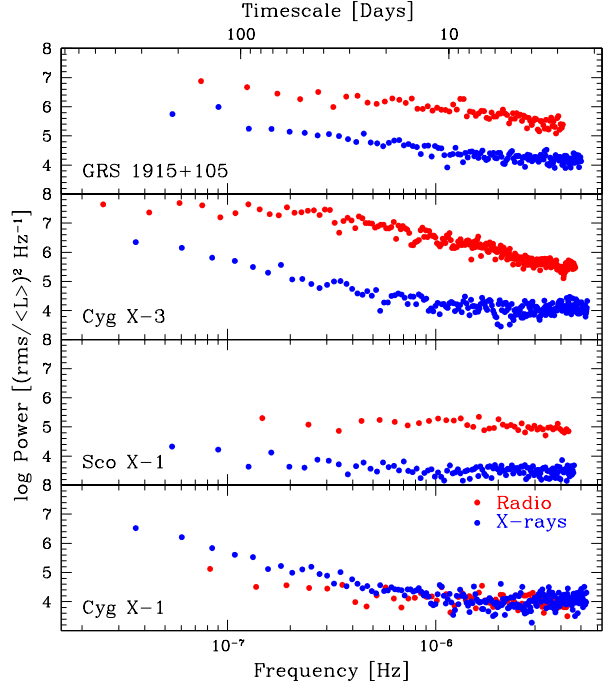}
\caption{Radio (2.25 ~GHz; red symbols) and X-ray (2-10 keV; blue
symbols) power spectra of GRS\,1915+105, Cyg\,X-3, Sco\,X-1 and
Cyg\,X-1.}  }
\end{figure}

\section{Long-term variability of microquasars in radio and in X-rays}

\begin{figure}
\centering{
\label{figsi}
\includegraphics[width=.6\textwidth]{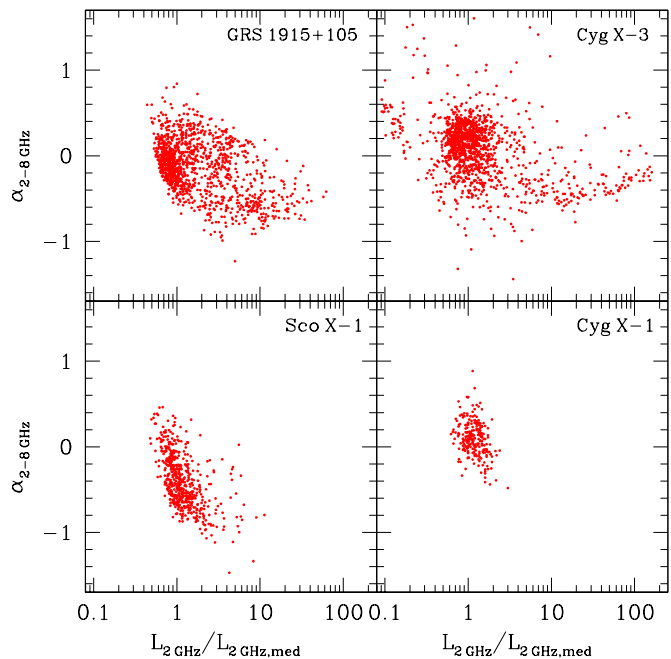}
\caption{Radio spectral index $\alphaghz$ as a function of the
  2.25~GHz radio luminosity in units of its median value. Each data
  point refers to the average over one day; only data points with
  $1-\sigma$ uncertainties on $\alphaghz$ smaller than 0.25 are
  plotted.} }
\end{figure}

Figure~\ref{figps} plots the power spectra of the four considered
microquasars in radio (at 2.25 GHz; red symbols) and X-rays (blue
symbols), in units of $(\rms/\av{L})^2 Hz^{-1}$, where $\av{L}$ is the
average luminosity.  It is apparent that the radio and X-ray power
spectra of each source have similar shape, but, while in the case of
Cyg~X-1 their normalisations are also similar, for the other three
sources the radio power spectra are higher (by factors up to $\sim30$)
than the corresponding X-ray power spectra. This reflects the fact
that, while the X-ray luminosity never exceeds three times its median
value, the radio luminosity is above this threshold for a significant
fraction of the time in GRS\,1915+105, Cyg\,X-3, and Sco\,X-1, with
peaks up to two orders of magnitude above the median. On the other
hand, in Cyg~X-1 the radio luminosity behaves in this respect as the
X-ray luminosity, being always below three times the median.  These
results suggest that two distinct modes contribute to radio activity:
the `coupled' mode, producing also X-ray emission, and the `flaring'
mode, producing strong radio flares, but no corresponding strong X-ray
emission.  This interpretation is supported by the diagrams in
Fig.~\ref{figsi}, showing, for daily-averaged data points of the four
sources, the radio spectral index
$\alphaghz\equiv\log(\Leightghz/\Ltwoghz)/\log(8.3/2.25)$ as a
function of $\Ltwoghz$ in units of its median value $\Ltwoghzmed$
(where $\Leightghz$ and $\Ltwoghz$ are the 8.3 GHz and 2.25 GHz
luminosities, respectively). In the case of GRS~1915+105, Cyg~X-3 and
Sco~X-1 there is an apparent segregation of the data points in these
diagrams: in the flaring mode (i.e., when $\Ltwoghz\gsim
3\Ltwoghzmed$) the radio spectrum is typically steeper than in the
coupled mode ($\Ltwoghz\lsim3\Ltwoghzmed$). Consistently, Cyg~X-1,
which has not been observed flaring during the radio monitoring
period, always has a flat radio spectrum.  GRS\,1915+105, Cyg X-3, Sco
X-1 are in the flaring mode 21, 10, and 3 per cent of the time,
respectively.  In terms of the canonical X-ray states of microquasars,
the coupled mode is associated with both the high/soft and the
low/hard states, while the flaring mode occurs at the transition
between these two states, in the so-called intermediate X-ray
state. Both the coupled and flaring mechanisms are active in the
neutron-star powered source Sco X-1, so neither mechanism is
associated with the extraction of energy from black-hole
spin. However, we note that Sco X-1 presents some difference with
respect to the three black-hole powered sources: it is less variable
in X-rays (see Fig.~\ref{figps}) ---probably because of the
contribution to the X-ray luminosity from the surface of the neutron
star--- and tends to have steeper radio spectrum also in the coupled
mode (see Fig.~\ref{figsi}).

\section{Implications for the origin of radio-loudness of quasars} 

Long-term radio and X-ray monitoring of microquasars indicates that
these sources, during their lifetime, often switch between a
radio-loud (flaring) and a radio-quiet (coupled) mode. If AGN behave
similarly, radio-loudness of quasars would be just a function of the
epoch at which the source is observed.  It must be noted that the
radio luminosity produced in the flaring mode is expected to be
proportionally smaller in microquasars than in radio galaxies because
the ambient medium around microquasars may not be dense enough to
produce bright lobes~(\cite{Hei02}). However, with an appropriate
choice of the threshold for radio-loudness (see \cite{NBB05}), the
fraction of time $\fmicroQ$ spent by a microquasar in the flaring mode
can be compared with the fraction of quasar $\fQ$ which are
radio-loud. On average we found $\fmicroQ\sim10\%$, and Ivezi{\' c} et
al.~(\cite{Ive02}) report $\fQ\sim8\%$ for the Sloan Digital Sky
Survey quasars.

The flaring emission has associated jets with relatively large bulk
Lorentz factors (e.g. \cite{Mil04}) and is typically characterised by
a steeper radio spectrum (see Fig.~\ref{figsi}) than the coupled
emission, so we speculate that in AGN the flaring mode may be the one
responsible for producing radio lobes, while the coupled mode produces
only core emission. Thus, different pieces of information suggest a
parallel between microquasars in the flaring mode and FR\,I or FR\,II
sources, and between microquasars in non-flaring states (either
low/hard or high/soft) and radio-quiet AGN (with only core radio
emission).  The absence of extended low-frequency radio lobes
associated with radio-quiet quasars is not inconsistent with this
picture, because the radiative lifetimes of synchrotron particles in
lobes are expected to be relatively short (\cite{BR00}).  A
consequence of the proposed scenario is that the radio luminosity
function of AGN will be dominated by the effects of variability: the
data from current radio surveys appear consistent with this hypothesis
(\cite{NB05}).

\end{document}